\documentclass[pra,onecolumn,showpacs]{revtex4}
\usepackage{amssymb}
\usepackage{amsmath}
\usepackage{graphicx}

\begin{document}

\title{Nonlinear management of topological solitons in a spin-orbit-coupled
system}
\author{Hidetsugu Sakaguchi$^{1}$ and Boris A. Malomed$^{2}$ }
\affiliation{$^{1}$Department of Applied Science for Electronics and Materials,
Interdisciplinary Graduate School of Engineering Sciences, Kyushu
University, Kasuga, Fukuoka 816-8580, Japan \\
$^{2}$Department of Physical Electronics, School of Electrical Engineering,
Faculty of Engineering, and Center for Light-Matter Interaction, Tel Aviv
University, Tel Aviv 69978, Israel}

\begin{abstract}
We consider possibilities to control dynamics of solitons of two types,
maintained by the combination of cubic attraction and spin-orbit coupling
(SOC) in a two-component system, namely, semi-dipoles (SDs) and mixed modes
(MMs), by making the relative strength of the cross-attraction, $\gamma $, a
function of time periodically oscillating around the critical value, $\gamma
=1$, which is an SD/MM stability boundary in the static system. The
structure of SDs is represented by the combination of a fundamental soliton
in one component and localized dipole mode in the other, while MMs combine
fundamental and dipole terms in each component. Systematic numerical
analysis reveals a finite bistability region for the SDs and MMs around $%
\gamma =1$, which does not exist in the absence of the periodic temporal
modulation (\textquotedblleft management"), as well as emergence of specific
instability troughs and stability tongues for the solitons of both types,
which may be explained as manifestations of resonances between the
time-periodic modulation and intrinsic modes of the solitons. The system can
be implemented in Bose-Einstein condensates, and emulated in nonlinear
optical waveguides.
\end{abstract}

\maketitle

\section{Introduction}

An active direction in the current work with Bose-Einstein condensates
(BECs) in atomic gases is using them as a testbed for emulation of various
effects originating in condensed-matter physics, which may be reproduced in
a clean and easy-to-control form in ultracold bosonic gases \cite%
{simulator,simulator2,simulator3}. In particular, a binary gas, with a
pseudo-spinor two-component wave function, may emulate spin-orbit coupling
(SOC) in semiconductors, i.e., the interaction between the electron's spin
and its motion across the underlying ionic lattice \cite{Dresselhaus,Rashba}%
, as first demonstrated in Ref. \cite{Nature}, see also reviews \cite{rf:1}-%
\cite{Zhai}. While most experimental works on the BEC\ simulation of SOC
dealt with effectively one-dimensional (1D) settings, implementation of SOC
in the quasi-2D geometry was reported too \cite{2D-experiment}, making it
relevant to consider 2D (and 3D) systems coupled by the spin-orbit
interaction. In this way, SOC opens a straightforward way to the creation of
topological modes characterized by vorticity, because linear operators
accounting for the coupling of two components in the corresponding system of
Gross-Pitaevskii equations (GPEs), see Eq. (\ref{GP}) below, generate
vorticity in one component if the other one is taken in the zero-vorticity
form.

The SOC effect, being linear by itself, may be naturally combined with the
intrinsic nonlinearity of bosonic gases, represented by cubic terms in the
respective GPEs \cite{Pit} (and/or by nonlocal cubic terms accounting for
long-range interactions in BEC built of dipole atoms \cite{dip-dip}). The
interplay of SOC and nonlinearity makes it possible to predict a great
variety of stable modes, including 1D and 2D solitons \cite%
{1D,2D,Chiquillo,EPL} and various nonlinear topological states in 2D, such
as vortices and vortex lattices \cite{Fukuoka}-\cite{Fukuoka10} and
skyrmions \cite{skyrmion}. In fact, the\ 2D\ and 3D SOC systems is one of
the most prolific sources of nonlinear states with intrinsic topological
structures.

A majority of works addressing nonlinear dynamics of SOC systems
investigated the case of self-repulsion, which is relevant to the current
experiments with $^{87}$Rb \cite{Nature}. Nevertheless, the interplay of SOC
with intrinsic attraction is possible too. Theoretical considerations
predict that the latter setting gives rise to 2D states with intrinsic
topological structures and very unusual dynamical properties: until
recently, it was commonly assumed that any 2D model with cubic
self-attraction may only generate unstable self-trapped states, such as
Townes solitons \cite{Townes} and their vortical counterparts \cite%
{Minsk1,Minsk2}. The fundamental (zero-vorticity) Townes solitons are
destabilized by the critical collapse (or by the supercritical collapse in
3D), while vortex solitons are subject to a still stronger splitting
instability \cite{review,review2}. A new paradigm was revealed by the
analysis of the 2D SOC system with cubic self- and cross-attractive
interactions \cite{Sakaguchi,Sherman1,Sherman2}: the linear SOC terms lift
the specific conformal invariance of the cubic GPEs in 2D, which is
responsible for the instability, as it makes norms of the entire soliton
family degenerate, allowing them to take a single value -- exactly the
critical one which launches the 2D collapse. As a result of lifting the
degeneracy, the norm of the 2D solitons falls \emph{below} the critical
value, thus protecting them against the onset of the collapse \cite%
{Sakaguchi}. The specific form of SOC creates two distinct species of
topological solitons in this case, namely, \textit{semi-vortices} (SVs),
which combine vorticities $S=0$ and $S=\pm 1$ in the two components, and
\textit{mixed modes} (MMs), which juxtapose zero-vorticity and vortex terms
in both components \cite{Sakaguchi}. Their stability is determined by
relative strength $\gamma $ of the cross-attraction between the components
and self-attraction (or the XPM/SPM (cross/self-phase modulation) ratio, in
terms of optics \cite{Maimist}): the SVs and MMs are stable (actually,
realizing the system's ground state) at $\gamma \leq 1$ and $\gamma \geq 1$,
respectively. The stability boundary shifts to $\gamma >1$ under the action
of the Zeeman splitting \cite{Sherman1}. On the other hand, SVs and MMs are
stable at all values of $\gamma $ in a model where the self-trapping is
provided not by attractive interactions, but by repulsion, with the local
strength growing fast enough from the center to periphery \cite%
{Frontiers,comment}; moreover, even excited states of SVs and MMs, which are
completely unstable in the case of the self-attraction \cite{Sakaguchi}, are
partly stable in the latter case.

SOC\ implemented in 2D BEC as the emulation of the solid-state phenomenology
may, in turn, be emulated in optical media, in terms of the spatiotemporal
propagation in dual-core planar waveguides, which simulates the
pseudo-spinor (two-component) structure of the wave field, while the SOC
proper is simulated by temporal \cite{optics} or spatial \cite{we} shift of
the linear coupling between the parallel waveguiding cores. The former
possibility is provided by the known effect of temporal dispersion of the
coupling \cite{Kip}, while the latter scheme may be supported by a skewed
structure of the medium separating the cores \cite{we}. Combining these
settings with the natural Kerr self-focusing in the dielectric material
opens the way to predict stable 2D optical solitons with an intrinsic
topological structure \cite{optics,we}, which may be construed as
counterparts of the above-mentioned SVs and MMs.

A known method which makes it possible to additionally stabilize soliton
modes which are, otherwise, vulnerable to instabilities, is the \textit{%
nonlinearity management}, i.e., periodic modulation of the self-focusing
strength in time, in the case of BEC \cite{Fatkh,Ueda,Kevr,VPG,breather,Itin}%
, or along the propagation distance in optics \cite{Isaac,book}. In the 2D\
SOC system, the application of the management technique is an especially
interesting possibility, as, by means of the Feshbach resonance controlled
by a low-frequency ac magnetic field \cite{Feshbach}, one can apply
time-periodic modulation to the above-mentioned XPM/SPM ratio $\gamma $. As
a result, the system will periodically pass from the SV-stability region, $%
\gamma <1$, to the MM-stability one, $\gamma >1$. Such a possibility poses
the problem of the existence and stability of the two species of the 2D
solitons in such dynamical states. The situation is somewhat similar to the
earlier studied situation with the Feshbach-resonance management
periodically alternating the sign of the nonlinearity in the
single-component case, which drives periodic transformations between regular
and gap-type solitons \cite{alternate}.

The present work addresses this dynamical problem by means of simulations of
GPEs including the SOC terms and the time-modulated coefficient $\gamma $.
However, performing such systematic simulations in the 2D model with many
control parameters is a challenging numerical problem. On the other hand, it
may be efficiently emulated by the similar 1D system, where a counterpart of
the SV is a \textit{semi-dipole} (SD), with a fundamental (spatially even)
structure in one component, and a dipole structure (a spatially odd
localized state with zero at the central point) in the other (see, e.g.,
Ref. \cite{Chiquillo}). MM states are possible in 1D as well, with the
fundamental and dipole terms mixed in both components. A crucially important
property of the 1D system, which suggests to use it for the emulation of the
2D prototype which is critically sensitive to the sign of $\gamma -1$, is
the fact that, exactly like in 2D, the one-dimensional SDs and MMs are
stable, respectively, at $\gamma <1$ and $\gamma >1$, and the
nonlinearity-controlling techniques, such as the Feshbach resonance, apply
even easier in the 1D settings.

The rest of the paper is organized as follows. The model is introduced in
Section II, along with some analytical results which help to illustrate the
structure of static SD\ and MM solitons. Results of the numerical analysis
of the management model are collected in Section III. In particular, a
region of the SD-MM bistability is found, and qualitative explanations are
presented for specific dynamical features (instability troughs and stability
tongues) induced by the management. The paper is concluded by Section IV.

\section{The model and analytical results}

In scaled units, the system of GPEs for the 1D binary BEC under the action
of the Rashba-type SOC and attractive nonlinearity, takes the form of \cite%
{Sherman1,Sherman2}
\begin{eqnarray}
i\frac{\partial \phi _{+}}{\partial t} &=&-\frac{1}{2}\frac{\partial
^{2}\phi _{+}}{\partial x^{2}}-(|\phi |_{+}^{2}+\gamma |\phi _{-}|^{2})\phi
_{+}+\lambda \frac{\partial \phi _{-}}{\partial x},  \notag \\
&&  \label{GP} \\
i\frac{\partial \phi _{-}}{\partial t} &=&-\frac{1}{2}\frac{\partial
^{2}\phi _{-}}{\partial x^{2}}-(|\phi |_{-}^{2}+\gamma |\phi _{+}|^{2})\phi
_{-}-\lambda \frac{\partial \phi _{+}}{\partial x},  \notag
\end{eqnarray}%
where $\phi _{+}$ and $\phi _{-}$ are two components of the pseudo-spinor
wave function, $\lambda $ is the SOC\ strength, self-attraction coefficients
are also scaled to be $1$, and $\gamma $ is the above-mentioned relative
strength of the nonlinear cross-interaction. By means of scaling
transformation,%
\begin{equation}
\phi _{\pm }=\lambda \tilde{\phi}_{\pm },~x=\lambda ^{-1}\tilde{x}%
,~t=\lambda ^{-2}\tilde{t},  \label{tilde}
\end{equation}%
which does not affect $\gamma $, we further fix $\lambda =1$ in Eq. (\ref{GP}%
) ($\lambda $ is kept as a free parameter in analytical expressions given
below by Eqs. (\ref{SDsmallN})-(\ref{exact}), to make the structure of those
expressions clearer). Of course, transformation (\ref{tilde}) cannot be
applied in the absence of SOC, $\lambda =0$.

Stationary solutions to Eq. (\ref{GP}) of the SD type, with chemical
potential $\mu <0$, have the form of
\begin{equation}
\phi _{+}^{(\mathrm{SD})}=e^{-i\mu t}f_{\mathrm{even}}(x),~\phi _{-}^{(%
\mathrm{SD})}=e^{-i\mu t}f_{\mathrm{odd}}(x),  \label{SD}
\end{equation}%
where $f_{\mathrm{even,odd}}(x)$ are even and odd functions, respectively
(see, e.g., Eq. (\ref{exact}) below). For the MMs, the appropriate ansatz is%
\begin{equation}
\phi _{\pm }^{(\mathrm{MM})}=e^{-i\mu t}\left[ f_{\mathrm{even}}(x)\pm f_{%
\mathrm{odd}}(x)\right] .~  \label{MM}
\end{equation}%
Stationary modes are characterized by their norm,%
\begin{equation}
N\equiv \int_{-\infty }^{+\infty }\left[ |\phi _{+}(x)|^{2}+|\phi
_{-}(x)|^{2}\right] dx.  \label{N}
\end{equation}

As mentioned above, for constant $\gamma =\gamma _{0}$, SDs and MMs are
stable, severally, at $\gamma _{0}<1$ and $\gamma _{0}>1$ (in Ref. \cite%
{Sakaguchi}, the latter result was originally obtained for SOC\ of the
Rashba type, but it was then established that the same is true for a general
Rashba-Dresselhaus combination \cite{Sherman1}). At $\gamma _{0}=1$, both
SDs and MMs are marginally stable.

We introduce the nonlinearity management by making $\gamma $ a periodic
function of time, the corresponding frequency, $\omega $, being a free
parameter of the management:%
\begin{equation}
\gamma (t)=\gamma _{0}-\gamma _{1}\sin \left( \omega t\right) .
\label{gamma}
\end{equation}%
Thus, in our model, with $\lambda =1$ fixed by the scaling, there are four
free parameters: $N$, $\gamma _{0}$, $\gamma _{1}$ and $\omega $. We report
systematic numerical results for $N=3$, checking that this value adequately
represents the generic case. Then, there remain three free parameters to
vary: $\gamma _{0}$, $\gamma _{1}$ and $\omega $ in Eq. (\ref{gamma}).

While the governing equations are cast here in the scaled form, the
rescaling can be undone to estimate characteristic values of control
parameters in physical units. In particular, in terms of the BEC realization
for light atoms, such as $^{7}$Li, and the assuming the size of the soliton $%
\sim 3$ $\mathrm{\mu }$m, the time unit in the scaled equations corresponds
to $t_{0}\sim 10$ ms. Then, characteristic scaled values of the management
frequencies, $\omega \sim 1$ (see below) correspond, roughly, to $\omega
\sim 2\pi \times 20$ Hz.

A more specific situation corresponds to the limit of $N\ll 1$, i.e., broad
small-amplitude solitons filled by the \textit{striped phase} \cite{striped}%
, which is represented below by factors $\cos \left( \lambda x\right) $ and $%
\sin \left( \lambda x\right) $ in Eqs. (\ref{SDsmallN}) and (\ref{MMsmallN}%
). In the absence of the management, they take the following approximate
forms, for the SD and MM species, respectively (here, for the clarity's
sake, we keep the SOC strength, $\lambda $, as a free parameter, rather than
fixing $\lambda =1$):%
\begin{eqnarray}
\phi _{+}^{(\mathrm{SD})}(x) &\approx &e^{-i\mu t}\frac{\sqrt{3+\gamma }}{4}%
N~\mathrm{sech}\left( \frac{3+\gamma }{8}Nx\right) \cos \left( \lambda
x\right) ,  \notag \\
&&  \label{SDsmallN} \\
\phi _{-}^{(\mathrm{SD})}(x) &\approx &-e^{-i\mu t}\frac{\sqrt{3+\gamma }}{4}%
N~\mathrm{sech}\left( \frac{3+\gamma }{8}Nx\right) \sin \left( \lambda
x\right) ,  \notag
\end{eqnarray}%
\begin{equation}
\phi _{\pm }^{(\mathrm{MM})}(x)\approx e^{-i\mu t}\frac{\sqrt{3+\gamma }}{4%
\sqrt{2}}N~\mathrm{sech}\left( \frac{3+\gamma }{8}Nx\right) \left[ \cos
\left( \lambda x\right) \pm \sin \left( \lambda x\right) \right] ,
\label{MMsmallN}
\end{equation}%
the chemical potential being $\mu \approx -(1/2)\left[ \lambda ^{2}+\left(
3+\gamma \right) ^{2}(N/8)^{2}\right] $ for both species ($\mu \leq -\lambda
^{2}/2$ is the semi-infinite spectral gap which may be populated by soliton
states).

Another specific case is one corresponding to large $N$, i.e., narrow
solitons. In particular, the respective SD state has a large fundamental
component,
\begin{equation}
\phi _{+}^{(\mathrm{SD})}\approx \exp \left( -i\frac{N^{2}}{8}t\right) \frac{%
N}{2}\mathrm{sech}\left( \frac{N}{2}x\right) ,  \label{largeN+}
\end{equation}%
while a relatively small dipole one,
\begin{equation}
\phi _{-}^{(\mathrm{SD})}=\exp \left( -i\frac{N^{2}}{8}t\right) u(y),
\label{largeN-}
\end{equation}%
with $y\equiv Nx/2$ and real function $u(y)$ determined by a linearized
equation,%
\begin{equation}
\left( \frac{1}{2}-\frac{1}{2}\frac{d^{2}}{dy^{2}}-\gamma~\mathrm{sech}%
^{2}y\right) u=-\lambda \frac{\sinh y}{\cosh ^{2}y}.  \label{u}
\end{equation}%
It is worthy to note that Eq. (\ref{u}) admits an \emph{exact solution}
precisely in the case of $\gamma=1$, when both SD and MM are stable:%
\begin{equation}
u(y)=-\frac{\lambda y}{\cosh y}.  \label{exact}
\end{equation}%
The application of the nonlinearity management to these specific cases
should be considered elsewhere.

\begin{figure}[h]
\begin{center}
{\normalsize \includegraphics[height=4.cm]{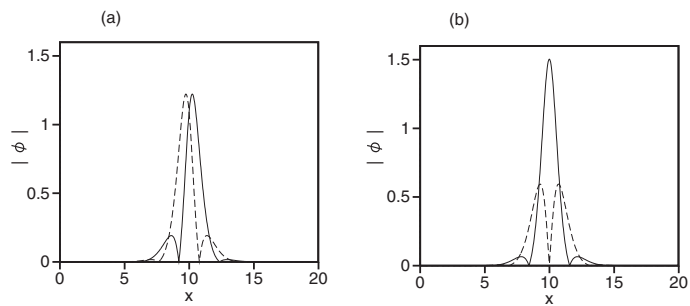} }
\end{center}
\caption{(a) The ground-state soliton of the MM (mixed-mode) type, produced
as a numerical solution of Eq. (\protect\ref{GP}) with $\protect\gamma %
_{1}=0 $ in Eq. (\protect\ref{gamma}), for $\protect\gamma _{0}=1.1$. (b) A
ground-state soliton of the SD\ (semi-dipole) type, for $\protect\gamma %
_{0}=0.9$. The solid and dashed lines show the profiles of $|\protect\phi %
_{+}(x)|$ and $|\protect\phi _{-}(x)|$ respectively.}
\label{fig1}
\end{figure}

Simulations of Eq. (\ref{GP}) were run starting with the initial state built
as a stable stationary soliton of the SD or MM type (the ground state),
produced by means of the imaginary-time-integration method applied to Eq. (%
\ref{GP}), with $\gamma _{1}=0$ in Eq. (\ref{gamma}). Figures \ref{fig1}(a)
and (b) display typical profiles of the corresponding static MM and SD
states, obtained at $\gamma _{0}=1.1$ and $\gamma _{0}=0.9$, respectively.
In the simulations of the full management model, with $\gamma _{1}\neq 0$ in
Eq. (\ref{gamma}) for various values of $\omega $, stable solitons were
identified as those which keep their integrity and initial structure (SD or
MM) in the course of long real-time simulations until $t=1000$. Note that,
in terms of the above-mentioned estimates for physical parameters of the BEC
setting, this corresponds to times $\gtrsim 10$ s, which definitely covers
the range of times that may be realized in any experiment.

\section{Results: stability regions for solitons under the action of the
management}

\begin{figure}[h]
\begin{center}
{\normalsize \includegraphics[height=4.cm]{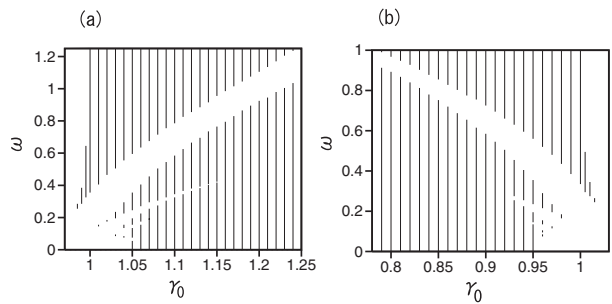} }
\end{center}
\caption{Shaded are stability regions in the parameter plane of $(\protect%
\gamma _{0},\protect\omega )$ for (a) MM- and (b) SD-type solitons at a
fixed management amplitude, $\protect\gamma _{1}=0.05$ in Eq. (\protect\ref%
{gamma}). Solitons are unstable in blank areas.}
\label{fig2}
\end{figure}

Systematic simulations of Eq. (\ref{GP}), with $\gamma $ taken as per Eq. (%
\ref{gamma}), are summarized in Fig. \ref{fig2}, which displays stability
regions in the parameter plane of $(\gamma _{0},\omega )$ for the MM- and
SD-type solitons, in panels (a) and (b), respectively, with a fixed value of
the management amplitude, $\gamma _{1}=0.05$ in Eq. (\ref{gamma}). One
conclusion is that the application of the management somewhat expands the MM
and SD\ stability areas, from the above-mentioned half-planes in the absence
of the management ($\gamma _{1}=0$), i.e., $\gamma _{0}\geq 1$ and $\gamma
_{0}\leq 1$, respectively, to values which may be smaller than $\gamma
_{0}=1 $ for the MMs, and larger than $1$ for the SDs. The expansion gives
rise to an MM-SD bistability region, which is presented in detail at the end
of this section.

\begin{figure}[h]
\begin{center}
{\normalsize \includegraphics[height=3.5cm]{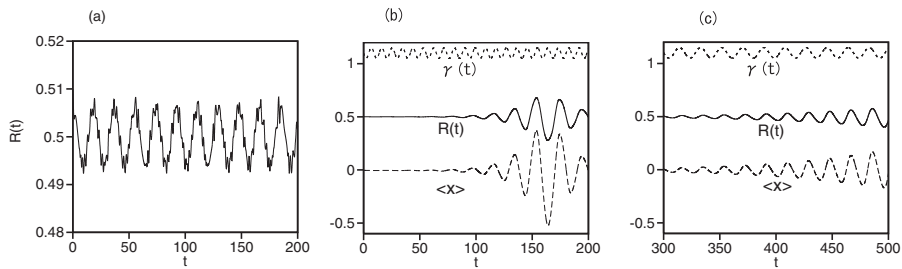} }
\end{center}
\caption{(a) The evolution of the norm ratio $R(t)$ (see Eq. (\protect\ref{R}%
)) for the MM, initially perturbed as per Eq. (\protect\ref{pert}), with $%
\protect\delta =0.01$, at $\protect\gamma _{0}=1.1$ and $\protect\gamma %
_{1}=0$. (b) The evolution of $R(t)$ (solid line) and $\langle x\rangle $
(dashed line) are shown, along with the underlying modulation format $%
\protect\gamma (t)$ (Eq. (\protect\ref{gamma}),\ the dotted line), for the
MM initiated by input (\protect\ref{pert}) with $\protect\delta =0.001$, at $%
\protect\gamma _{0}=1.1$, $\protect\gamma _{1}=0.05$, and $\protect\omega %
=0.68$. (c) The same as in (b), but for $\protect\omega =0.335$.}
\label{fig3}
\end{figure}

Conspicuous features revealed by Fig. \ref{fig2} are instability troughs
created by the management in the originally stable half-planes. These
features may be understood as manifestations of resonant interaction of the
ac drive, supplied by the management, and an eigenmode of intrinsic
oscillations of stable solitons, existing in the absence of the management.
To consider this possibility, the MM's intrinsic eigenmode and respective
eigenfrequency were identified by direct real-time simulations of Eq. (\ref%
{GP}), with $\gamma _{1}=0$ in Eq. (\ref{gamma}), adding a small initial
perturbation to the numerically exact MM soliton, as follows:
\begin{equation}
\phi _{+}(x;t=0)=\left( 1+\delta \right) \phi _{0+},~\phi _{-}(x;t=0)=\left(
1-\delta \right) \phi _{0-},  \label{pert}
\end{equation}%
where $\left\{ \phi _{0+},\phi _{0-}\right\} $ are the components of the MM
state displayed in Fig. \ref{fig1}(a) (i.e., with $\gamma _{0}=1.1$). For
the perturbation strength $\delta =0.01$ in Eq. (\ref{pert}), the ensuing
evolution of an essential characteristic of the perturbed soliton, which we
define as the share of the total norm staying in one component,
\begin{equation}
R(t)=N^{-1}\int_{-\infty }^{+\infty }|\phi _{+}(x)|^{2}dx,  \label{R}
\end{equation}%
is presented in Fig. \ref{fig3}(a). The evolution of $R(t)$ clearly exhibits
eigenfrequency $\omega _{0}\approx 0.34$ of the MM's intrinsic mode.

The most conspicuous feature in Fig. \ref{fig2}(a) is a relatively wide
diagonal instability trough. In particular, at $\gamma _{0}=1.1$ its size is
\begin{equation}
0.585<\omega <0.785.  \label{width}
\end{equation}%
To consider a possible explanation of this feature in terms of the
resonance, in Fig. \ref{fig3}(b) we display a typical example of the
instability for $\gamma _{1}=0.05$ at the point taken in the middle of
interval (\ref{width}),%
\begin{equation}
\omega /2=0.34\approx \omega _{0}.  \label{parres}
\end{equation}%
The instability is shown by means of the corresponding time dependence for $%
R(t)$ and the MM's center-of-mass position,%
\begin{equation}
\langle x\rangle =N^{-1}\int_{-\infty }^{+\infty }\left[ |\phi _{+}\left(
x\right) |^{2}+|\phi _{-}\left( x\right) |^{2}\right] xdx,  \label{x}
\end{equation}%
together with the modulation format, $\gamma (t)$. The initial condition is
taken as Eq. (\ref{pert}) with $\delta =0.001$. The instability manifests
itself, in Fig. \ref{fig3}(b), by oscillations of $R(t)$ and $\langle
x\rangle $ with a growing amplitude (the spontaneously emerging oscillatory
motion of unstable solitons is illustrated by an example displayed below in
Fig. \ref{fig5} (d)). Figure \ref{fig3}(b) shows the initial stage of the
perturbation growth. At larger $t$, $R(t)$ exhibits irregular oscillations
with an amplitude $\simeq 0.22$ around $R=0.5$. The situation observed in
Fig. \ref{fig3}(b) is a typical picture of mechanical instability caused by
the parametric resonance with the frequency ratio $1:2$ \cite{LL}, in
agreement with relation (\ref{parres}).

Further, Fig. \ref{fig3}(c) shows the same dynamical characteristics, $R(t)$%
, $\langle x\rangle $ and $\gamma (t)$ for $\gamma _{0}=1.1$, $\gamma
_{1}=0.05$, and $\omega =0.335$, which is a point belonging to the second
(much more narrow) instability trough in Fig. \ref{fig2}(a). In this case,
the instability is again manifested by the growth of $R(t)$ and $\langle
x\rangle $, although the instability is weaker than in Fig. \ref{fig3}(b).
Comparing the current value of $\omega $ with $\omega _{0}$ (see Eq. (\ref%
{parres})), we conclude that this instability may be interpreted as caused
by a direct resonance, with frequency ratio $1:1$. A plausible explanation
of additional small \textquotedblleft notches" in Fig. \ref{fig2}(a) is the
presence of very weak higher-order (subharmonic) resonances. Similarly, the
instability-trough pattern observed in Fig. \ref{fig2}(a) can be construed
as manifestations of the parametric, direct, and subharmonic resonances with
an intrinsic mode of the SD soliton.

\begin{figure}[tbp]
\begin{center}
{\normalsize \includegraphics[height=3.5cm]{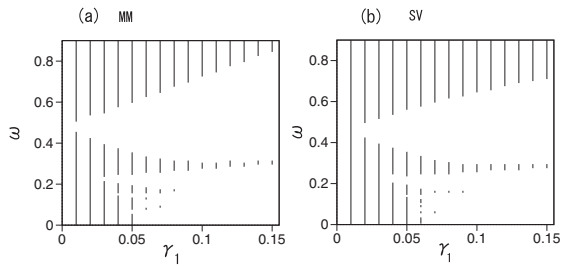} }
\end{center}
\caption{Shaded are stability regions in the parameter plane of the
management parameters, $(\protect\gamma _{1},\protect\omega )$ (see Eq. (%
\protect\ref{gamma})) for MM-type solitons at fixed $\protect\gamma %
_{0}=1.05 $ (a) and SD states (b) at $\protect\gamma _{0}=0.95$.}
\label{fig4}
\end{figure}

Another relevant summary of the stability results is presented in Fig. \ref%
{fig4} for the solitons of the MM (a) and SD (b) types in parameter plane $%
(\gamma _{1},\omega )$, fixing the value of the constant term in the
nonlinearity coefficient (\ref{gamma}), \textit{viz}., $\gamma _{0}=1.05>1$
in (a), and $\gamma _{0}=0.95<1$ in (b). The solitons are stable at
sufficiently large $\omega $, where the high-frequency modulation is
effectively averaged out, hence it does not produce a conspicuous effect. It
is also natural that stability areas tend to shrink as $\gamma _{1}$
increases. Nevertheless, narrow stability tongues are found too. For
example, the SD soliton is stable at $0.265<\omega <0.295$ for $\gamma
_{1}=0.12$. Examples of the SD dynamical states, taken at the same fixed
value of the management amplitude, $\gamma _{1}=0.12$, are shown in Fig. \ref%
{fig5}(b) for (a) $\omega =0.31$ (unstable), (b) $\omega =0.28$ (stable),
and (c) $\omega =0.25$ (unstable). The stable SD in panel (b) belongs to the
narrow \textquotedblleft tongue" in Fig. \ref{fig4}(b). It was found that
the instability region above the \textquotedblleft tongue" is similar to the
main instability trough shown in Fig.~\ref{fig2}(a).
\begin{figure}[h]
\begin{center}
{\normalsize \includegraphics[height=3.5cm]{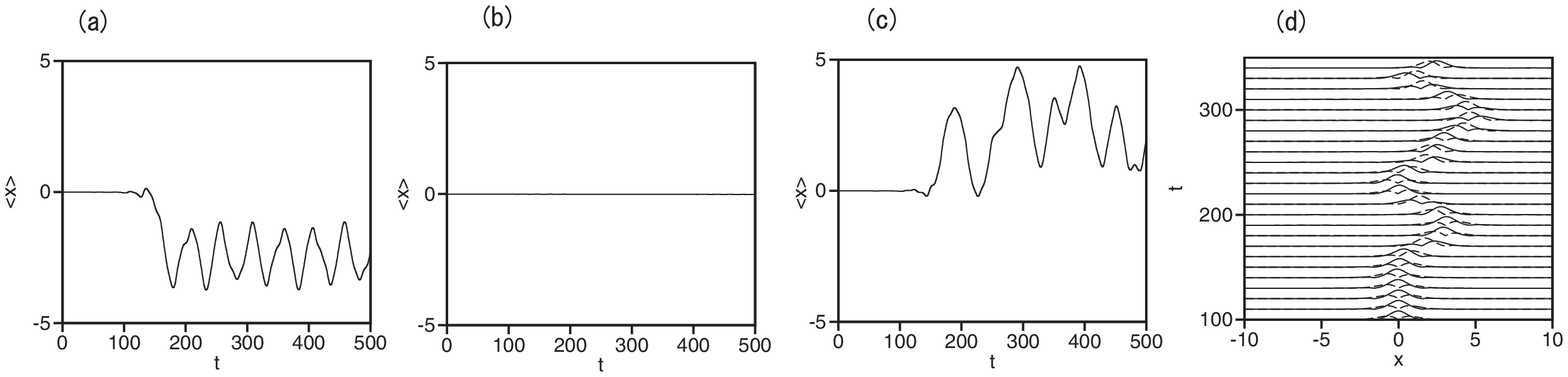} }
\end{center}
\caption{The evolution of the center-of-mass position, $\langle x\rangle $,
for the SD solitons, shown along with the underlying time-periodic
modulation $\protect\gamma (t)$, with $\protect\gamma _{0}=0.95$ and $%
\protect\gamma _{1}=0.12$ in Eq. (\protect\ref{gamma}), for three values of
the modulation frequency: (a) $\protect\omega =0.31$, (b) $\protect\omega %
=0.28$, and (c) $\protect\omega =0.25$. Panel (d) illustrates the unstable
evolution of the soliton corresponding to (c).}
\label{fig5}
\end{figure}

While accurate explanation of the nature of the stability tongues requires a
more detailed analysis, it is plausible that they also originate from a
nonlinear resonance, which, as it is known, gives rise to both unstable and
stable solution branches \cite{LL}.

\begin{figure}[h]
\begin{center}
{\normalsize \includegraphics[height=3.5cm]{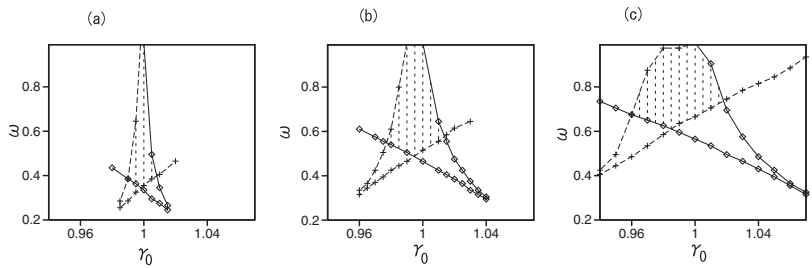} }
\end{center}
\caption{Stability boundaries for the SD and MM dynamical states (solid and
dashed lines, respectively) at $\protect\gamma _{1}=0.05$ (a), $\protect%
\gamma _{1}=0.1$ (b), and $\protect\gamma _{1}=0.15$ (c). The bistability
holds between the boundaries, in the vertically-shaded areas (see the text).}
\label{fig6}
\end{figure}

Lastly, as mentioned above, a noteworthy feature observed in Fig. \ref{fig2}
is bistable coexistence of the SD and MM states in a small but finite region
near $\gamma _{0}=1$ (in the absence of the management, $\gamma _{1}=0$, the
bistability is only possible strictly at $\gamma _{0}=1$ \cite{Sakaguchi}).
Figure \ref{fig6} shows stability boundaries for the SD solitons (solid
lines) and MMs (dashed lines) in the parameter space of $(\gamma _{0},\omega
)$ (the same plane which is displayed in Fig. \ref{fig2}), at three fixed
values of the management amplitude: (a) $\gamma _{1}=0.05$; (b) $\gamma
_{1}=0.1$; (c) $\gamma _{1}=0.15$. The SD and MM solitons are stable,
severally, in regions bounded by solid lines and dashed ones. Accordingly,
the bistability takes place in finite areas between the dashed and solid
lines, which are designated by vertical shading in Fig. \ref{fig6}. The
bistability region is small, but its size increases with the growth of $%
\gamma _{1}$.

\section{Conclusion}

In this work, we address the possibility to control stability and dynamics
of two types of two-component spin-orbit-coupled solitons, SDs
(semi-dipoles) and MMs (mixed-modes), which represent distinct species of 1D
topological modes, by means of the nonlinearity management, i.e., making the
relative strength of the cross-attraction, $\gamma $, a function of time
which periodically oscillates around $\gamma =1$. This value is the boundary
between stability regions of the static SDs and MMs. By means of systematic
simulations, we have found a finite bistability region around $\gamma =1$,
which expands with the increase of the management amplitude. In the usual
SDs' and MMs' stability domains ($\gamma <1$ and $\gamma >1$, respectively),
the analysis reveals new features generated by the management, in the form
of long instability troughs and, on the other hand, stability tongues
penetrating into instability domains. These features may be explained as
manifestations of resonances between the time-periodic management and
excitation modes of stable 2D solitons, that exist in the absence of the
management.

As said above, the 1D soliton species of the SD and MM types emulate their
2D spin-orbit-coupled counterparts, in the form of SVs (semi-vortices) and
two-dimensional MMs, respectively. It may be quite interesting, although
somewhat challenging, in terms of collecting systematical numerical data, to
identify stability charts for the 2D topological modes of both types under
the action of the nonlinearity management. Another relevant extension may be
realization of the nonlinearity management for the two-component system
combining SOC and long-range dipole-dipole interactions \cite{Guangzhou}. It
may be realized by periodically varying the direction of the external
magnetic field which determines the orientation of the atomic magnetic
moments.

\section*{Acknowledgments}

The work of B.A.M. is supported, in part, by the Israel Science Foundation
through grant. No. 1287/17, and that of H.S. by a Grant-in-Aid for
Scientific Research (No. 18K03462) from the Ministry of Education, Culture,
Sports, Science and Technology of Japan.

\end{document}